\documentclass{jpp}
\usepackage{graphicx}
\usepackage{epstopdf, epsfig,amsmath}

 \newcommand{\bfE}{\mathbf{E}}
\newcommand{\bE}{\mathbf{E}}
\newcommand{\bfB}{\mathbf{B}}
 
\newcommand{\bB}{\mathbf{B}}
\newcommand{\bfJ}{\mathbf{J}}
\newcommand{\bJ}{\mathbf{J}}
\newcommand{\bfv}{\mathbf{v}}

\newcommand{\bfx}{\mathbf{x}}
\newcommand{\bv}{\mathbf{v}}

\shorttitle{Guidelines for authors}
\shortauthor{G. Lapenta, D. Gonzalez-Herrero, E. Boella }

\title{Multiple scale kinetic simulations with the energy conserving semi implicit particle in cell (PIC) method}

\author{Giovanni Lapenta\aff{1}
  \corresp{\email{valsusa@gmail.com}},
  Diego Gonzalez-Herrero\aff{1}
 \and Elisabetta Boella\aff{1}}

\affiliation{\aff{1}Department of Mathematics, KU Leuven, University of Leuven, Belgium}

\begin{document}

\maketitle

\begin{abstract}
The recently developed energy conserving semi-implicit method (ECsim) for PIC simulation  is applied to multiple scale problems where the electron-scale physics needs to be only partially retained and the interest is on the macroscopic or ion-scale processes. Unlike hybrid methods, the ECsim is  capable of providing kinetic electron information, such as wave-electron interaction (Landau damping or cyclotron resonance) and non-Maxwellian electron velocity distributions. However, like hybrid, the ECsim does not need to resolve all electron scales, allowing time steps and grid spacing orders of magnitude larger than in explicit PIC schemes. The additional advantage of the ECsim is that the stability at large scale is obtained while conserving energy exactly.  Three examples are presented: ion acoustic waves, electron acoustic instability and reconnection processes.
\end{abstract}

\section{Introduction}


Kinetic models are based on the description of the velocity distribution function for electrons and ions. An often-used approach is to describe this distribution statistically, employing a sample of particles: this is the PIC method~\citep{birdsall-langdon,hockney-eastwood}. A critical distinction within PIC methods is between explicit and implicit algorithms. Explicit algorithms alternatively move particles (in the fields known until that time) and advance the fields (with the sources provided by the particle velocity and position known until that time). This alternative advancing decouples the field and particle equations resulting in a very simple and highly computational efficient technique. However, the explicit method is limited in what ranges of time steps and grid spacing it can use: the electron scales need to be resolved down to the smallest scales \citep{birdsall-langdon,hockney-eastwood}. 

If the electron physics does not necessarily need to be resolved with great accuracy, the explicit method is not appropriate because the attempt to run the simulation with less resolution is met first with an excessive numerical heating due to the failure of energy conservation, followed by numerical instability if the resolution is too coarse \citep{birdsall-langdon,hockney-eastwood}. 

This limitation is eliminated by the implicit methods \citep{markidis2011energy,Chen-JCP-2011}, where particles and fields are advanced together in a non-linear iterative scheme. To remove the complexity of the non linear coupling  and to reduce the computational cost per time step, semi-implicit methods have been invented \citep{Brackbill:1985}. In this case, the coupling between particles and fields is approximated linearly and a self-consistent equation is solved for the fields alone without needing to iterate with the particles. Two previous semi-implicit classes of algorithms have been published: the direct implicit method (DIM) \citep{directimplicit} based on  expressing the particle response to field changes via a sensitivity matrix, and the implicit moment method (IMM) \citep{brackbill-forslund} based on approximating the plasma response using moments of the distribution function (typically up to the pressure tensor). 

While implicit methods conserve energy exactly, semi-implicit methods tend to either lose or gain energy depending on the specifics of the configuration of the run \citep{directimplicit-optimisation}. This lack of energy conservation results in limiting the range of resolution accessible. For example, in reconnection simulations \citep{lapenta2012particle} such as the one reported as an example here, the grid resolution needs to resolve the electron skin depth to represent the physics correctly, but it does not need to resolve the Debye length. Similarly, the particle trajectories need to be well resolved, a goal reached by resolving well the electron cyclotron motion, but the plasma frequency does not need to be resolved accurately. For  low density, low temperature plasmas, there is a vast gap between the  Debye and skin depth scale. In these situations, the semi-implicit methods struggle to conserve energy at a sufficiently good degree to prevent numerical instability.

A new semi-implicit method has been recently developed to conserve energy exactly, to machine precision \citep{lapenta2016exactly}. The first published tests confirm exact energy conservation and suggest that when the method is applied to multiple scale problems, the range of resolution where the method remains stable is wider than in previous semi-implicit schemes. 

This possible beneficial effect is put to test in the present paper showing that for three examples of multiple scale problems indeed the new method allows to cover a wide range of resolution without losing stability or energy conservation.  First, we consider the ion acoustic wave, where both electron and ion physics play a role in the evolution and the method shows its ability to run resolving only the ions scales without artificially heating the electrons. Then, the electron acoustic instability is analysed. Here, the concurrent presence of three species: ions and cold and hot electrons, whose dynamics is characterised by different temporal and spatial scales, makes the problem multi-scale. In this case, we show that ECsim is able to capture the correct physics even when the smallest scales are not resolved. Finally, the well known problem of the so-called GEM challenge \citep{birnGEM} is used to show that the only scales which need to be properly resolved are the electron skin-depth in space and the cyclotron period in time (instead of the electron Debye length and the inverse of the electron frequency). Stepping over the smaller electron scales (Debye and plasma scales) does not lead to loss of energy conservation or stability even for cold rarefied plasmas where these scales are orders of magnitude smaller than the resolved scales.



\section{Summary of the Energy Conserving Semi-Implicit PIC}
Recently a new semi-implicit PIC method  has been proposed to conserve energy exactly to machine precision \citep{lapenta2016exactly}.   The fundamental enabling step that led to the new method is a new mover that allows the explicit analytical calculation of the current generated by the particles during one time step without any approximation. 

\subsection{Particle Mover}
The new mover combines the DIM $D_1$ scheme \citep{Hewett:1987} with the IMM and ECPIC $\theta$-scheme \citep{brackbill-forslund}. The particle position is advanced as in the $D_1$ scheme, but the velocity is advanced as in  the $\theta$ scheme:
\begin{equation}
\begin{array}{c}
\displaystyle \bfx_p^{n+1/2}=\bfx_p^{n-1/2} + \Delta t \bfv_p^{n},\\ \\
\displaystyle \bfv_p^{n+1}=\bfv_p^{n} + \frac{q_p \Delta t}{m_p} \left(  \bfE^{n+\theta}(\bfx_p^{n+1/2}) + \overline{\bfv}_p\times \bfB^{n}(\bfx_p^{n+1/2}) \right),
\end{array}
\label{thetaECsim}
\end{equation}
where $\overline{\bfv}_p= (\bfv_p^{n+1}+ \bfv_p^{n})/2$.

 We evaluate the fields at the time staggered particle position $\bfx_p^{n+1/2}$ known explicitly from the time staggering of the leap-frog, but we use the implicit electric field at the $\theta$ time level: $\bfE^{n+\theta}$  as in the standard IMM. 
 The  fields are computed at the known position $\bfx_p^{n+1/2}$ rather than at the unknown position $\overline{\bfx}_p$ that in the standard IMM requires the predictor-corrector iteration.
These two positions are conceptually similar, expressing the particle position at the mid-time between the old and new evaluations of the velocity. But one is computed explicitly while the other is computed as part of a predictor-corrector iteration. Both are second order accurate, but the scheme in Eq.\,(\ref{thetaECsim}) is simpler to compute. The combined scheme is second order accurate  and has the same stability properties of the IMM.  This property allows one to write the scheme to be exactly energy conserving.

The equation for the velocity can be solved analytically to express explicitly $\overline{\bfv}_p$~\citep{vu}. Using vector manipulation, the velocity equation can be rewritten in the equivalent  form: 
\begin{equation}
\overline{\bfv}_p=\widehat{\bv}_p+
\beta_s\widehat{\bE}_p,
\label{theta-rotated}
\end{equation}
where hatted quantities have been rotated by the magnetic field:
\begin{equation}
\begin{array}{c}
 \widehat{\bv}_p = {\alpha}^n_p  \bv^n_p, \\ \\
\widehat{\bE}_p = {\alpha}^n_p, 
\bE_p^{n+\theta} 
\end{array}
\label{hatted}
\end{equation}
via a rotation matrix ${\alpha}_p^n$ 
defined as:
\begin{equation}
{\alpha}_p^n =  \frac{1}{1+(\beta_s B_p^{n})^2}
\left(\mathbb{I}-\beta_s \mathbb{I} \times \bB_p^n +\beta_s^2
\bB_p^n \bB_p^n \right),
\label{alpha}
\end{equation}
where $\mathbb{I}$ is the dyadic tensor (matrix with diagonal of 1) and $\beta_s=q_p \Delta t/2m_p$ (independent of the particle weight
and unique to a given species). 

The fields at the particle positions are computed by interpolation:
\begin{align}
\bfE_p^{n+\theta}= \sum_g \bfE_g^{n+\theta} W(\bfx_p^{n+1/2}-\bfx_g),\\
\bfB_p^{n}= \sum_g \bfB_g^{n} W(\bfx_p^{n+1/2}-\bfx_g),
\label{interp}
\end{align}
where we have assumed that the field equations are discretized on a grid with a generic index $g$. For brevity we introduced the notation: $\bfB_p^{n}=\bfB^{n+\theta}(\bfx_p^{n+1/2})$ and $\bfE_p^{n+\theta}=\bfE^{n+\theta}(\bfx_p^{n+1/2})$.
In the examples below, the interpolation functions $W$ are  b-splines of order $\ell=1$ \citep{bspline}.

\subsection{Field Solver}
For the Maxwell's equations we use the same discretization as in the standard IMM \citep{ipic3d}.  The two curl Maxwell's equations are discretized in time with another $\theta$-scheme:
\begin{equation}
\begin{array}{ccc}
\displaystyle \nabla_g \times \bfE^{n+\theta} + \frac{1}{c} \frac{\bfB^{n+1}_g-\bfB^n_g}{\Delta t} =0,\\ \\
\displaystyle \nabla_g \times \bfB^{n+\theta} - \frac{1}{c} \frac{\bfE^{n+1}_g-\bfE^n_g}{\Delta t} =\frac{4\pi}{c} \overline{\bfJ}_g,
\end{array}
\label{maxwell-discrete}
\end{equation}
where $\overline{\bfJ}_g$ is computed at the mid temporal location. For each species we use
\begin{equation}
\overline{\bfJ}_{sg}=\frac{1}{V_g}\sum_p q_p \overline{\bfv}_p W(\bfx_p^{n+1/2}-\bfx_g),
\label{currentECsim}
\end{equation}
where the summation is over the particles of the same species, labeled by $s$. The total current is obtained summing over the species.

The spatial operators in Eq. (\ref{maxwell-discrete}) are discretized on a grid, indicating $\nabla_g$ as a shorthand for the spatial discretization of the operators. In the  examples below  the same discretization of iPic3D is used~\citep{sulsky1991numerical,ipic3d,lapenta2012particle}, however, all the derivations below are not critically dependent on which spatial discretization is used.

\subsection{Current evaluation}
The set of Maxwell's and Newton's equations are coupled. In the spirit of the semi-implicit method, we do not want to solve two coupled sets with a single non-linear iteration and find instead a way to extract analytically from the equations of motion the information needed for computing the current without first moving the particles. In previous semi-implicit methods this is done via a linearization procedure. The new mover used here allows us, instead, to derive the current rigorously without any approximation. 

Substituting then Eq.\ (\ref{theta-rotated}) into Eq.\ (\ref{currentECsim}), we obtain without any approximation or linearization:
 \begin{equation}
\overline{\bfJ}_{sg}=\widehat{\bJ}_{sg}+\frac{\beta_s}{V_g}\sum_p q_p {\alpha}^{n}_p
\bE_p^{n+\theta}  W_{pg},
\label{currentECsim2}
\end{equation}
where we shortened the notation $W_{pg}=W(\bfx_p^{n+1/2}-\bfx_g)$  and the summation is intended over all particles of species $s$.

The following hatted currents were defined:
\begin{equation}
\widehat{\bJ}_{sg} = \sum_p q_p  \widehat{\bv}_p W_{pg},
\label{hattedmoments}
\end{equation}
representing the current based on the  hatted velocities. 

Computing then the electric field on the particles by interpolation from the grid as in Eq.\,(\ref{interp}), Eq.\,(\ref{currentECsim2}) becomes:
\begin{equation}
\overline{\bfJ}_{sg}=\widehat{\bJ}_{sg}+\frac{\beta_s }{V_g}\sum_p \sum_{g^\prime} q_p {\alpha}^{n}_p
 \bE_{g^\prime}^{n+\theta}  W_{pg^\prime} W_{pg}.
\label{currentECsim3}
\end{equation}

The formula above is conveniently expressed introducing mass  matrices~\citep{burgess1992mass} defined by elements as
\begin{equation}
M_{s,gg^\prime}^{ij} = \sum_p q_p {\alpha}^{ij,n}_p W_{pg^\prime} W_{pg}.
\label{mass-matrices}
\end{equation}
There are  3$v$  such matrices, where $v$ is the dimensionality of the magnetic field and velocity vector, not to be confused with the dimensionality of the geometry used for space $d$. The indices $i$ and $j$ in Eq. (\ref{mass-matrices}) vary in the 3$v$-space. For example for full 3-components vectors, $i,j=1,2,3$ and there are 9 mass matrices.  Each matrix is symmetric and very sparse with just $2d$ diagonals.  In matrix notation the $3v$ mass matrices defined above are written as $M_{gg^\prime}$, i.e.  without the indices $i,j$ for the vector directions. 

Using the mass matrices, the current becomes:
\begin{equation}
\overline{\bfJ}_{sg}=\widehat{\bJ}_{sg}+\frac{\beta_s }{V_g}\sum_{g^\prime} M_{s,gg^\prime}
 \bE_{g^\prime}^{n+\theta}.
\label{currentECsimfinal}
\end{equation}

Equation (\ref{currentECsimfinal}) is the central ingredient of the ECsim method: it expresses the  advanced current at the mid-point of the time step and the electric field at the advanced time. This linear relationship can be substituted into the discretized Maxwell's equations (\ref{maxwell-discrete}) to form a linear set of equations to be solved on the grid:
\begin{equation}
\left\{ \begin{array}{l}
\displaystyle \nabla_g \times \bfE^{n+\theta} + \frac{1}{c} \frac{\bfB^{n+1}-\bfB^n}{\Delta t} =0,\\ \\
\displaystyle \nabla_g \times \bfB^{n+\theta} - \frac{1}{c} \frac{\bfE^{n+1}-\bfE^n}{\Delta t} =\frac{4\pi}{c} \left( \widehat{\bJ}_{g}+\sum_{g^\prime} M_{gg^\prime}
 \bE_{g^\prime}^{n+\theta} \right),
\end{array}
\right.
\label{mexwellECIMM}
\end{equation}
where we have introduced the total current $\widehat{\bJ}_{g} = \sum_s \widehat{\bJ}_{sg}$ and the species summed mass matrices that written by elements are:
\begin{equation}
M_{gg^\prime}^{ij} =\sum_s\frac{\beta_s }{V_g}M_{s,gg^\prime}^{ij},
\end{equation}
which can be more usefully held in memory, reducing the memory consumption by a factor equal to the number of species.
\subsection{Energy Conservation} \label{th_energy_conservation}

The method described above, in the specific case $\theta=1/2$, satisfies an exact energy conservation principle proven in \citet{lapenta2016exactly}: 
\begin{equation}
\begin{array}{l}
\displaystyle\sum_g \frac{ (\bfB_g^{n+1})^2-(\bfB_g^n)^2 }{4 \pi}+ \sum_g\frac{ (\bfE_g^{n+1})^2-(\bfE_g^n)^2 }{4\pi} = \\ \\
\displaystyle \Delta t \sum_g\overline{\bfJ}_g \cdot \overline{\bfE}_g +
\frac{c\Delta t}{4 \pi}\sum_g \nabla_g \cdot  (\bfE_g \times \bfB_g ).
\end{array}
\label{energy-fields}
\end{equation}

The  left hand side is recognized as the variation of the magnetic and electric energy. The first term on the right hand side is the energy exchange term with the particles and the last term is the divergence of the Poynting flux.  This is the usual  equation for electromagnetic energy conservation. 

The examples below will further confirm in practice the validity of the result above, energy is conserved to machine precision. 

 

 \section{Massively parallel implementation}
%
%
%
%
%
The new  ECsim  has been implemented in a parallel code, which was
built over the very basic structure of the implicit moment PIC code
iPic3D \citep{ipic3d}. Despite the differences between the algorithm used in iPic3D and
the one presented here, there are many modules of iPic3D
that could be retained in the new code, like the particle communications between
processors or the input/output procedure. As its predecessor,  the new code has
been written in C and C++ and it uses MPI for parallel communication between
processes. 

Similarly to iPic3D, the new code uses a three dimensional Cartesian grid to
compute the fields where the particles are immersed. When several
processors are used, the physical domain is divided into sub-domains. In order to
compute the derivatives, each process not only owns one sub domain, but also  the
first cells of its neighbours (ghost cells). Those ghost values are communicated
between processes through MPI routines. Analogously, each process owns the
particles, which belong to its domain. After every cycle, the particles that
have left the sub-domain are communicated to the right process via MPI routines.
The results of the simulation are written on disk using the HDF5 
format for fields and particles, and in ASCII for
other additional data (as the energy of the system or the input data used).
Except for that, everything is different in the new code: the moment gathering,
the field solver and the particle mover.

In order to solve the field equations (Eq. \eqref{mexwellECIMM}),  the
implicit current and the mass matrices need to be computed in advance. The
calculation of the implicit current is different from what was done in
iPic3D. Here, for each particle it is necessary to interpolate the  magnetic
field from the grid to the particle position, then compute the $\alpha$ matrix
and once the implicit current for that particle has been calculated, interpolate
it to the grid. The mass matrices were not present in iPic3D, and their
calculation is the most time demanding part of the code.  For each
node there are 27 three-by-three mass matrices ($g'$ can take the value of $g$
and all the neighbors nodes). However, due to their symmetry 
($M_{gp} \equiv M_{pg}$), only 14 matrices must  be stored in memory for each
node, which means that 126 scalar values have to be computed for each node. 

In the ECsim algorithm both the electric and the magnetic field are solved
together whereas in iPic3D first the electric field is obtained and then the
magnetic field is computed as the curl of the electric field. Thus 
the linear system is twice as big as the one in
iPic3D. For this reason we decided to use PETSc
\citep{petsc-web-page} for the task.
PETSc is a suite of libraries, which provide several tools to deal with problems ruled by
differential equations. The main advantages of PETSc are that it is specially
intended for parallel calculations, meaning that it should scale well when
using many cores, and it allows the user to change between 
several linear and non linear solvers very easily. 

Finally, once the fields are known, the position and the velocity of the
particles can be updated. Unlike what happens in iPic3D, here the mover
has an explicit scheme (no iterations are needed). However, for each particle
the $\alpha$ matrix needs to be computed. Note that this matrix was calculated in
the moment gathering, but due to the high number of particles usually employed
in the simulations, it is not practical to store these matrices in memory. Instead,
the $\alpha$ matrix is computed again for each particle. This
implies  the interpolation of both the electric and magnetic fields from the 
grid to the particles. Despite this fact, the particle
mover in this code is much simpler than it was in iPic3D and it is less time
consuming.

\begin{figure}
  \centerline{\includegraphics[height=7cm]{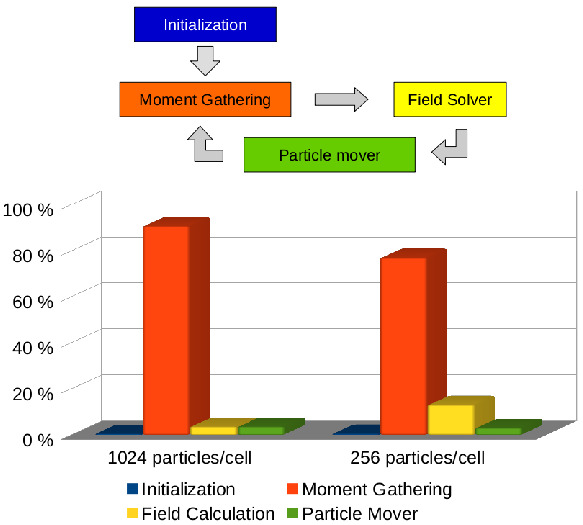}}
  \caption{Main loop of the ECsim and the percentage of the total time spent
          in each task in two cases: with 1024 and 256 particles per cell.}
\label{fig:4.1}
\end{figure}

In Fig. \ref{fig:4.1} the scheme of the main loop and the percentage of
the total time spent in each task are shown. The data correspond to two different simulations of magnetic reconnection. In both cases, the domain was discretised in 256x128 cells and the time step considered was  $\Delta t = 0.1 \, \omega_{pi}^{-1}$.  A different number of particles per cell has been employed in the two simulations: 256 and 1024. Both tests have been carried out with 16 MPI processes in a single node with an Intel® Xeon®
Sandy Bridge E5-2680 processor. The most time
demanding portion of the code is the moment gathering; in particular the calculation 
of the mass matrices is clearly the dominant part. As the number of particles decreases, 
we would expect the field solver time to become more important, however, even 
with only 256 particles per cell, the moment gathering is still the most time 
consuming part of the code. In all the tests performed, the time required by the particle 
mover is always negligible when compared  with the moment gathering. Taking into account that, 
the optimization of the code should be focused on the moment gathering stage, and in
particular on the mass matrix calculation. For instance, the vectorization of
the mass matrix calculation (the field interpolation between the grid and the
particles and the calculation of the elements of the mass matrices) will dramatically
improve the performance. Future work will address this aspect.

Regarding the computation time required by the new code, if we analyze the time
per cycle (with the same input values) ECsim is more time demanding than
iPic3D (it takes about 3 times more). However, if we look at the big picture,
taking into account the fact that in the new code the time step can be chosen 
much larger than in iPic3D, the new code will need less cycles to cover the same physical
time than iPic3D, and hence the total time of the simulation will be considerably 
reduced. Moreover, the grid size (i.e. the number of cells) is
no longer a constraint. This means that if we are not interested in the smallest
scales of the problem, we can use larger cells (that is, less cells) and then
reduce the time needed for each cycle.


  
 \section{Results}
 
 \subsection{Ion Acoustic Wave}
\begin{figure}
\begin{center}
\includegraphics[scale=0.25]{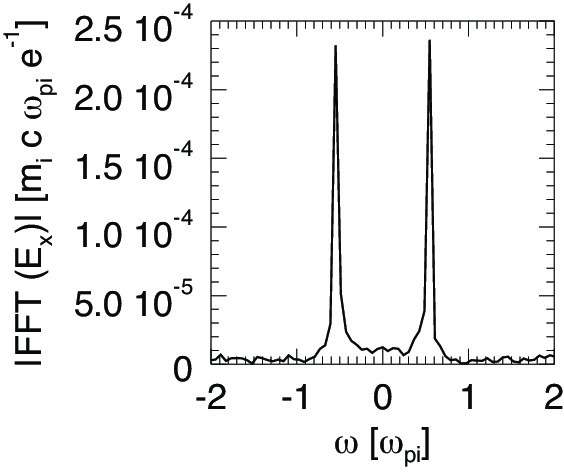}
\end{center}
\caption{Frequency domain Fourier transform of the electric field component corresponding to $k = 2 \pi/L = 45 \, \omega_{pi}/c$. The peaks are located at $\omega = \pm 0.5 \, \omega_{pi}$, in perfect agreement with the solution of Eq.~\eqref{dispersion_th}.}\label{Dispersion}
\end{figure} 

Ion acoustic waves are longitudinal low frequency modes, where both the ion and the electron dynamics play a role \citep{Gary_book}. Therefore, they represent an optimum multi-scale test for the ECsim algorithm. The wave is triggered by a density perturbation in a plasma composed of hot electrons and relatively colder ions. Considering Maxwellian electrons and ions with temperature $T_e$ and $T_i \ll T_e$ respectively, the wave dispersion relation is \citep{Gary_book}
\begin{equation}
\epsilon \left(\omega, k \right) = 1+\chi_e \left(\omega, k \right) +\chi_i \left(\omega, k \right) \label{dispersion_th},
\end{equation}
with the susceptibilities $\chi_j \left(\omega, k \right)$ given by
\begin{equation}
\chi_j\left(\omega, k \right) = \frac{\omega_{pj}^2}{k^2 v_{th,j}^2}\left[1+ \frac{\omega}{\sqrt{2} k v_{th,j}} Z \left(\frac{\omega}{\sqrt{2}k v_{th,j}} \right) \right], \label{suscept}
\end{equation}
where the subscript $j$ has been used to indicate the $j-th$ species, $\omega \ll \omega_{pe}$ and $k$ are the wave frequency and the wavenumber respectively, $\omega_{pj}=\sqrt{4\pi e^2 n_j/m_j}$ is the plasma frequency of the $j-th$ species having density $n_j$ and mass $m_j$, $e$ is the elementary charge, $v_{th,j}=\sqrt{T_j/m_j}$ is the thermal speed and $Z$ is the plasma dispersion function ~\citep{Fried_and_Conte}.

\begin{figure}
\begin{center}
\includegraphics[scale=0.25]{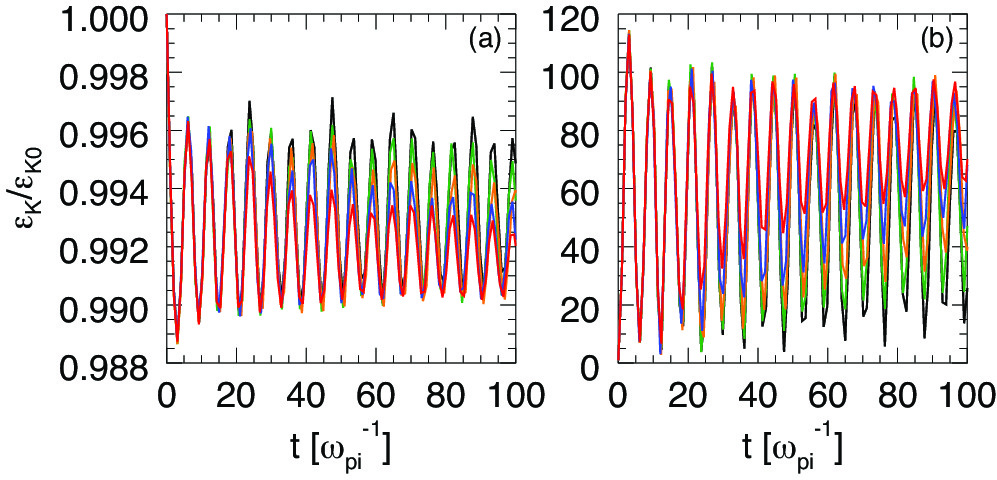}
\end{center}
\caption{Evolution of the electron (a) and ion (b) kinetic energy for $\Delta t =$ 0.0043 (black), 0.0089 (green), 0.01 (orange), 0.0133 (blue) and 0.0177 (red) $\omega_{pi}^{-1}$.}\label{IAW_kinetic_energy}
\end{figure} 

A series of numerical simulations were performed to check the code stability and accuracy. Similar parameters to those in \citep{Chen-JCP-2011} have been used. A homogeneous plasma composed of electrons and ions with a reduced mass-to-charge ratio of 200 was introduced. Hot electrons with $T_e = 20 \, \text{KeV}$ and ions with $T_i = 2 \, \text{eV}$ were considered. At $t=0$, a small sinusoidal perturbation was superimposed to the equilibrium density $n_0$:
\begin{equation}
n_j (t=0) = n_0 \left[1+0.2\cos\left(\frac{2\pi}{L}x\right) \right],
\end{equation} 
where $L = 0.14 \, c/\omega_{pi} $ is the simulation box length, $c$ is the speed of light in vacuum, $\omega_{pi}=\sqrt{4 \pi e^2 n_0/m_i}$ is the ion plasma frequency and $x$ is the longitudinal coordinate. The simulation box was discretised using 32 cells, so that  $\Delta x \simeq 0.3 \, \lambda_D$, with $\lambda_D=\sqrt{T_e/4 \pi e^2 n_0}$ electron Debye length. All the simulations used 32000 particles per species, which were pushed until $t_{end} \simeq 100 \, \omega_{pi}^{-1}$.

At first, code results have been compared with theory. In this case, a time step $\Delta t = 0.0043 \, \omega_{pi}^{-1}$ was employed. It is important to notice that this time step corresponds to the maximum value that can be used in an explicit electromagnetic PIC algorithm for stability reasons ($\Delta t_{\text{explicit}}< \Delta x/c$). Figure \ref{Dispersion} shows the Fourier transform in the frequency domain of the electric field. The field component corresponding to the excited mode $k =2\pi/L=45 \, \omega_{pi}/c$ is displayed. The peaks correspond to a wave frequency $\omega = \pm 0.5 \, \omega_{pi}$, in perfect agreement with the solution of Eq. \eqref{dispersion_th}. A detailed study was performed increasing progressively the time step until the semi-implicit limit $\Delta t_{\text{implicit}} < \Delta x/v_{th,e}$. Results are summarised in Fig. \ref{IAW_kinetic_energy}, where the evolution of the ion and electron kinetic energy is shown. In all the case considered here, the long term behaviour of the ion acoustic wave is well described. A stronger damping of the wave is observed when increasing the $\Delta t$ ~\citep{brackbill-forslund}, but results are overall convergent and the physics of the wave is captured correctly, even when a big $\Delta t$ is chosen. In all the simulations, the total energy is conserved down to round-off precision as demonstrated in Sect. \ref{th_energy_conservation}.


 \subsection{Electron Acoustic Instability}
\begin{figure}
\begin{center}
\includegraphics[scale=0.25]{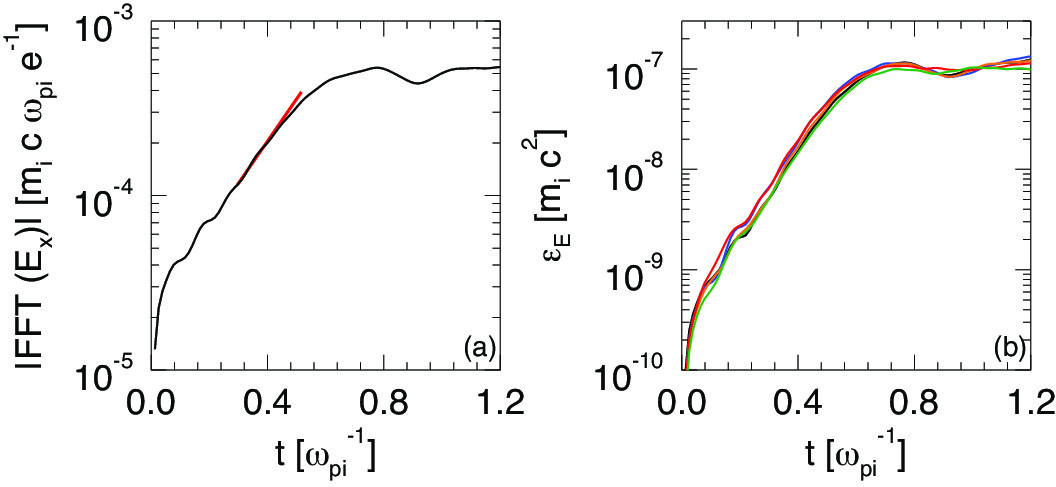}
\end{center}
\caption{(a) Evolution of the electric field component corresponding to $k=153 \, \omega_{pi}/c$ (black). The over-plotted red line represents the theoretical growth rate as provided by the solution of Eq. \eqref{disp_eaw}. (b) Evolution of the electric field energy for $\Delta x \simeq $ 1 (black), 2 (blue), 3.5 (orange), 7 (red) and 14 (green) $\lambda_{Dc}$. }\label{eaw_gr}
\end{figure} 

In a plasma where ions are at rest and two different populations of electrons with similar density, but different temperature are streaming against each other, when the relative drift is greater than the thermal velocity of the colder species, the electron acoustic instability may arise. The dispersion relation of the instability is given by \citep{Gary_book}
\begin{equation}
\epsilon \left(\omega, k \right) = 1+\chi_c \left(\omega, k \right) + \chi_h \left(\omega, k \right) +\chi_i \left(\omega, k \right) \label{disp_eaw},
\end{equation}
where the subscripts $c$, $h$ and $i$ indicate cold electrons, hot electrons and ions, respectively. If Maxwellian particles are considered, the ion susceptibility can be expressed by Eq.~\eqref{suscept}, while the electron susceptibilities are
\begin{equation}
\chi_j\left(\omega, k \right) = \frac{\omega_{pj}^2}{k^2 v_{th,j}^2}\left[1+ \frac{\omega+k v_{0,j}}{\sqrt{2} k v_{th,j}} Z \left(\frac{\omega+k v_{0,j}}{\sqrt{2}k v_{th,j}} \right) \right],
\end{equation}
being $j=c, \, h$ and $v_{0,j}$ the drift velocity.

Since $\lambda_{Dc} \ll \lambda_{Dh}$ and $\Gamma_{\max} \ll \omega_{ph} \ll \omega_{pc}$, where $\Gamma_{\max} $ is the maximum growth rate of the instability, the electron acoustic instability represents a good multi-scale test for the ECsim algorithm. A simulation box with length $L=0.334 \, c/\omega_{pi}$ filled with a homogenous three species plasma has been considered. Ions with realistic charge to mass ratio are distributed according to a Maxwellian with zero average speed and temperature $T_i=0.1 \, \text{KeV}$. The cold electrons with density $n_c = 0.8 \, n_i$ are characterised by bulk velocity $v_{0,c}= -0.069 \, c$ and temperature $T_c=T_i$. The density and the speed of the hot electrons are such that $n_c+n_h=n_i$ and $n_cv_{0,c}+n_hv_{0,h} = 0$, while their temperature is $T_h=10 \, \text{KeV}$. The instability is triggered by an initial density perturbation $\propto \sin \left [ k(\Gamma_{\text{max}})x \right ] $, where $\Gamma_\text{max} = 5.6 \, \omega_{pi}$ is given by the solution of Eq. \eqref{disp_eaw} and $k(\Gamma_{\text{max}}) = 153 \, \omega_{pi}/c$ is the corresponding wavenumber.

Code results have been compared with the numerical solution of Eq. \eqref{disp_eaw}. In this case the domain was discretised with 1024 cells, so that $\Delta x = 3.3 \times 10^{-4} \, c/\omega_{pi} \simeq \lambda_{Dc}$, being $\lambda_{Dc}$ the smallest spatial scale in the system. The time step was chosen to be the maximum time step allowed in an explicit code: $\Delta t \simeq \Delta x/c = 3.2 \times 10^{-4} \, \omega_{pi}^{-1}$. Figure \ref{eaw_gr}\,(a) shows the growth rate $\Gamma$ of the field component corresponding to $k = 153 \, \omega_{pi}/c$. The simulation growth rate is measured to be $\Gamma = 5.6 \, \omega_{pi}$, as predicted by the theory. 

\begin{figure}
\begin{center}
\includegraphics[scale=0.3]{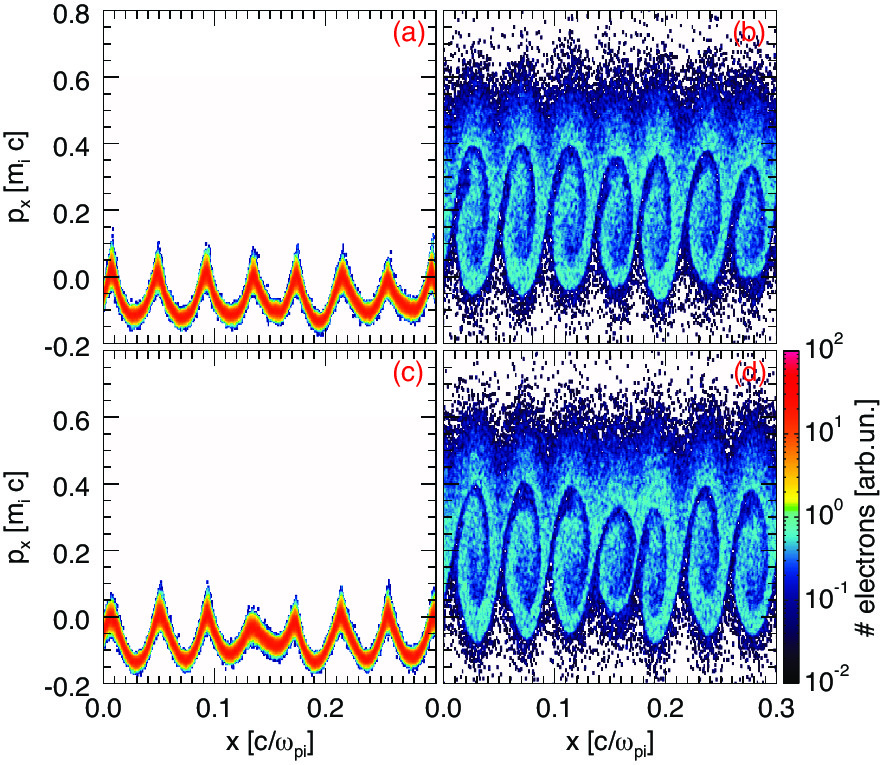}
\end{center}
\caption{Cold (a, c) and hot (b, d) electron phase space at the saturation of the instability ($t = 0.7 \, \omega_{pi}^{-1}$). Results (a) and (b) have been obtained with $\Delta x \simeq \lambda_{Dc}$, while (c) and (d) with $\Delta x \simeq 7 \, \lambda_{Dc}$.}\label{eaw_phasespace}
\end{figure}

A set of simulations have been carried out to check the code convergence and stability versus coarser spatial discretisation. Results are reported in Fig. \ref{eaw_gr}\,(b), which shows the evolution of the electric field energy for $\Delta x =  3.3 \times 10^{-4} \div 5.2 \times 10^{-3} \, c/\omega_{pi} \simeq 1\div 15 \, \lambda_{Dc}$. In all the cases analysed, the field energy shows the same trend. The correct growth rate of the instability can be retrieved by all the simulations and the same saturation level is reached. The particle phase spaces have been also compared, to determine whether kinetic effects were properly retained. Figure \ref{eaw_phasespace} shows the longitudinal phase space for cold and hot electrons at the saturation of the instability ($t = 0.7 \, \omega_{pi}^{-1}$), obtained with the finest and the coarsest resolution. Plots do not show appreciable differences: particle trapping is well described even when $\Delta x \gg \lambda_{Dc}$, confirming that results are accurate and the algorithm can reproduce well the physics with a coarse discretisation.  

  \subsection{Reconnection Example}
%
%
%
%
%


Magnetic reconnection \citep{biskamp} is an important example of electromagnetic processes with multiple scales. In the process of reconnection, electrons and ions become separated from the field lines, braking the frozen-in condition. In the process, magnetic energy is released to particle energy (macroscopic flows and heating). The critical aspect of interest here is that the reconnection process presents two scales: the ion become decoupled from the frozen-in flow on a scale comparable with the ion inertial length, $d_i=c/\omega_{pi}$, the electrons instead decouple in a much smaller scale, the electron skin depth, $d_e=c/\omega_{pe}$ \citep{birn-priest}. The Debye length does not play a significant role in this specific process, although it does play a role in other aspects especially along the separatrices \citep{Lapenta2010,lapenta2011bipolar,Divin2012,lapenta2014separatrices}.
For very cold plasmas the ratio of $d_e/\lambda_{De}$ can be very large. Explicit but even previous semi-implicit method are limited in how far they can exceed the condition $\Delta x/\lambda_{De}<\pi$. We show here one example, based on the classic GEM-challenge~\citep{birnGEM}.

The system is periodic in all directions and the
initial state corresponds to a double Harris equilibrium. In particular the
initial magnetic field in the x-direction is given by \citep{drake2005production}:
\begin{equation}
  \label{EQ4.2.01}
  B_x(y) = B_0 \left(-1 + \tanh \left(\frac{y - y_B}{\delta} \right) 
                                   + \tanh \left(\frac{y_T - y}{\delta} \right)
                          \right),
\end{equation}
where $L_x$ and $L_y$ are the dimensions of the box, $y_B=0.25 \, L_y$ and
$y_T=0.75 \, L_y$ the position of the two current sheets. The magnetic
field $B_0$ is computed from the pressure balance condition $n_0 (T_e + T_i) =
B_0^2/8\pi$, $T_e$ and $T_i$ being the temperature of the electrons and ions.
The magnetic field in z and y directions is zero as are all the components
of the electric field. All the particles of the same species are at the same thermal temperature with Maxwellian distribution and the drift velocity is set to ensure force
balance.

To trigger the nonstationary reconnection we add a
perturbation to the initial magnetic field. The perturbation chosen in this case
is the same as in \citep{Lapenta2010} and its vector potential is given by
\begin{eqnarray}
  \label{EQ4.2.02}
  \delta \mathbf{A}(x,y) &=& A_0 \cos(2 \pi \, (x-x_T) / \delta) \cos (\pi \, (y-y_T) / \delta)
               \, \mathrm{e}^{-((x-x_T)^2 + (y-y_T)^2)/\delta} + \\ \nonumber
                    && A_0 \cos(2 \pi \, (x-x_B) / \delta) \cos (\pi \, (y-y_B) / \delta)
               \, \mathrm{e}^{-((x-x_B)^2 + (y-y_B)^2)/\delta} \\ \nonumber,
\end{eqnarray}
where $x_B = 0.25 \, L_x$ and $x_T = 0.75 \, L_x$. The initial charge density of the 
system is given by Eq. \eqref{EQ4.2.03}
\begin{equation}
  \label{EQ4.2.03}
  \rho(y) = n_0 \left( \mathrm{sech}^2 (y - y_T)  + \mathrm{sech}^2(y - y_B) \right) +
              n_{\infty},
\end{equation}
where $n_0$ is the density of particles with drift velocity and $n_\infty$
is the density of the background particles (particles without drift velocity). This perturbation is simular to that in the standard GEM challenge~\citep{birnGEM} but more localized in space
in \citep{Lapenta2010}.  The perturbation is strong bringing reconnection immediately to a non-linear state, avoiding the need for a long linear phase of slow growth. This is the central feature of the GEM challenge that makes it a widely used benchmark for new codes.

We have carried out two sets of simulations, in both we use $m_i/m_e = 25$,
$T_i/T_e=5$, $n_0/n_{\infty} = 1$, $\delta=0.5$, $L_x=25$ and $L_y=12.5$ (where
the length of the box is in units of the skin depth of the ions). The number of
cells in each direction are  $256 \times 128 \times 1$. In the case of high temperature
the thermal velocity of the electrons is  $v_{\mathrm{th}e}=0.1 $ which leads to $B_0=0.0693$ 
and in the case of low temperature $v_{\mathrm{th}e} = 0.001$  with $B_0=000693$ (the velocity 
is given in units of the speed of light and $B_0$ in units of the Alfven magnetic 
field). 

For the high temperature case we use two values of the time step $\Delta t =
0.1$ and $\Delta t = 0.4$ (in units of the inverse of the ions plasma frequency). 
In the simulations performed with iPic3D the damping parameter used is $\theta = 1$ ,
otherwise the energy dramatically increases and the code crashes. With ECsim we
use $\theta = 0.5$ (in which case the energy is exactly conserved) and $\theta =
1.0$, which damps out the high frequencies and hence prevents energy conservation.
It is important to note that conversely to what happens in iPic3D,  when the energy is 
not conserved in ECsim, it is because some frequencies are damped, which means that
the energy associated to them is lost. Therefore, the energy will always decrease 
and the system is stable.

\begin{figure}
  \centerline{\includegraphics[height=6cm]{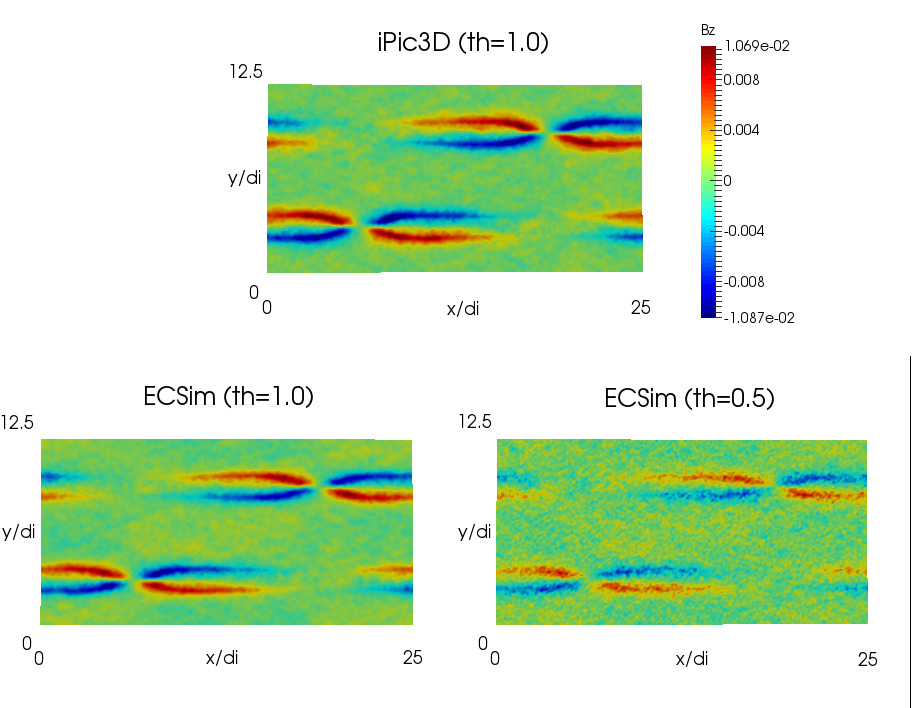}}
  \caption{High temperature. Cases with $\Delta t=0.4 \, \omega_{pi}^{-1}$, out of plane magnetic
           field ($B_z$) in $t = 240 \, \omega_{pi}^{-1}$.}
\label{fig:4.2.1}
\end{figure}

In Fig. \ref{fig:4.2.1} we see the magnetic field out of plane (z-component)
from three different simulations, iPic3D ($\theta = 1$), ECsim with $\theta = 1$
and ECsim with $\theta = 0.5$. All the results are in good agreement and the 
only difference is that in the case with $\theta = 0.5$ there is more noise, 
which is due to the high frequencies which were damped in the other cases.
\begin{figure}
  \centerline{\includegraphics[height=4cm]{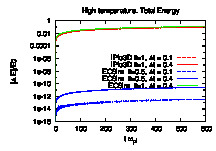}}
  \caption{High temperature. Absolute value of the variation of the total energy 
          (in logarithmic scale) for different values of the time step.}
\label{fig:4.2.2}
\end{figure}

\begin{figure}
  \centerline{\includegraphics[height=4cm]{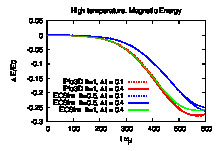}}
  \caption{High temperature. Variation of the magnetic energy for different 
           values of the time step.}
\label{fig:4.2.3}
\end{figure}
In Fig. \ref{fig:4.2.2} the variation of the total energy is shown and, as we
expected, the total energy is conserved up to the tolerance of the field solver
(in all cases $10^{-12}$) in the results from ECsim  with $\theta = 0.5$. The
evolution of the magnetic energy (Fig. \ref{fig:4.2.3}), which gives us an idea
of the reconnection rate, is very similar in all cases. 

\begin{figure}
  \centerline{\includegraphics[height=4cm]{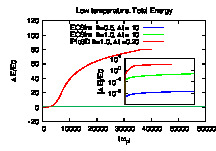}}
  \caption{Low temperature. Variation of the total energy in linear
           scale and its absolute value in logarithmic scale (inner figure).}
\label{fig:4.2.4}
\end{figure}

\begin{figure}
  \centerline{\includegraphics[height=4cm]{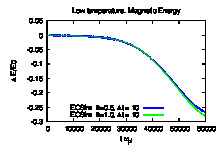}}
  \caption{Low temperature. Variation of the magnetic energy.}
\label{fig:4.2.5}
\end{figure}

The situation in the low temperature case is completely different. If in the
high temperature case we needed to simulate up to $t \approx 600 \, \omega_{pi}^{-1}$,
here we will need $t \approx 60000 \, \omega_{pi}^{-1}$. If we use the same time step
the simulation would be 100 times more expensive, and even in that case, iPic3D
is not capable of dealing with this situation: after a given number of cycles
the energy dramatically increases (see Fig. \ref{fig:4.2.4}) and the
simulation has no physical meaning. On the other hand, in ECsim the energy is
exactly conserved and hence we are able to use a larger time step without any
problems. But even with $\theta = 1$ (which no longer guarantees energy 
conservation), the energy remains stable, it changes: but the variation is small
enough to not distort the results. This can be seen in Fig.
\ref{fig:4.2.5} where the variation of the magnetic energy from ECsim with 
$\theta = 1$ and $\theta = 0.5$ is shown. 

\begin{figure}
  \centerline{\includegraphics[height=6cm]{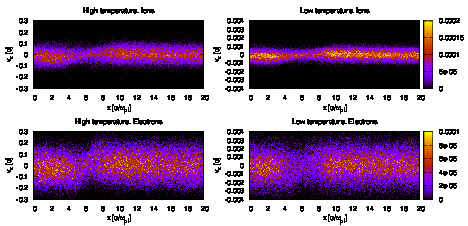}}
  \caption{Phase space for the x component of the velocity as a function of the x direction of the particles
           located in the band $3.0 \; d_i < y < 3.1 \; d_i$. In color the charge density is shown. The scale
           of the ion charge density is shown in the right top side and the one of the electron in the right bottom
           part.}
\label{fig:4.2.6}
\end{figure}

\begin{figure}
  \centerline{\includegraphics[height=6cm]{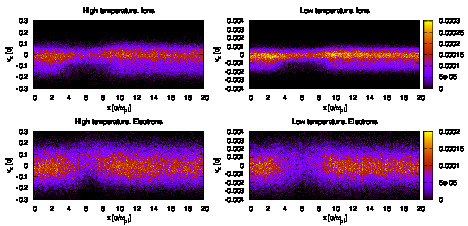}}
  \caption{Phase space for the z component of the velocity as a function of the x direction of the particles
           located in the band $3.0 \; d_i < y < 3.1 \; d_i$. In color the charge density is shown. The scale
           of the ion charge density is shown in the right top side and the one of the electron in the right bottom
           part.}
\label{fig:4.2.7}
\end{figure}

The new method can capture the  kinetic level of description for both species. While some aspects of reconnection can be captured accurately by fluid models (see \citet{birn-priest} for a review), to obtain the phase space distribution of electron and ions, full kinetic description is needed. As reconnection develops, the average electron flows determine the electric and magnetic field structure. The most typical feature is the formation of the Hall magnetic field, seen in Fig. \ref{fig:4.2.1}. These effects can be captured also by fluid models. But the results reported in Fig.~\ref{fig:4.2.6} and \ref{fig:4.2.7} are typical of kinetic approaches. 

In Fig. \ref{fig:4.2.6}, the electron and ion phase space is reported in the cross section $(x, v_x)$. The four panels compare ions and electrons and the cold and hot case (note the different velocity axis). Ions are accelerated at the reconnection site (located near $x/d_i=6$). We are showing here the distribution function for particles located in the range $y/d_i=[3,3.1]$ and are integrated in the two velocities not reported. 
Figure \ref{fig:4.2.7} reports the distribution function for the same $y$ range but in the plane  $(x, v_z)$ (integrated in the other two velocities). 

For the distribution in $v_x$, at each $x$ the distribution is fairly Maxwellian, as can be observed by its symmetry with respect with the peak. This effect is due to the nature of the particle acceleration along $x$ that is caused by the Lorentz force associated with the vertical magnetic field~\citep{Goldman:2011}.  Instead, the distribution changes strongly in space, being much narrower at the reconnection site and broadening  away from it. 

The distribution in $v_z$, is highly non-Maxwellian. The acceleration along the $z$-direction is caused by the reconnection electric field active effectively in the electron diffusion region where the electrons become demagnetized and instead of $\bfE \times \bfB$ drifting are accelerated~\citep{moses1993plasma,divin2010model}. A consequence if this mechanism is the strong distortion of the particle distribution in the vicinity of the x-point. Moving away from the x-point the distribution remains strongly non-Maxwellian, especially for the ions, with a distinct asymmetry between positive and negative $v_z$, the negative side having a much wider width. 

In summary, we have shown that, in the situations where iPic3D can be used, the
new code gives the same result, and in those set-ups in which iPic3D is
limited, the new code is able to give good results, and even more, to speed up
the simulation time by using a larger time step. The new method is capable not only 
of capturing the non-linear macroscopic effects of reconnection, such as the production of the Hall magnetic field but also the detail microphysics of particle acceleration at the kinetic level. 



\section{Conclusions}
We presented the ECsim algorithm and tested its ability to handle multiple scales. The ECsim is based on two innovations. First, a new mover is presented, a hybrid of the $\theta$-scheme and the leap-frog algorithm. The new mover is still implicit and unconditionally stable but it allows to more readily compute the interpolation between particles and cells. Second, we use a new method to compute the current needed for Maxwell equations. The new method introduces a number of mass matrices that produce an exact representation of the current. The relationship between the electric field and the current mediated by the mass matrix is an exact consequence of the mover without any linearization and it is naturally linear. 
These two innovations lead to one critical consequence: the new ECsim is exactly energy conserving. This point is important in two aspects. First, energy conservation is a desirable property because energy is of course conserved in reality and missing this aspect, as most PIC method do, is unsatisfactory. Second, the existence of energy conservation is in itself a proof on non-linear stability in $L_2$-norm. The fact that an energy integral exist to limit the energy error, eliminating the tendency of PIC methods to numerically heat. 

Compared with explicit PIC \citep{birdsall-langdon}, the new method eliminates all instabilities and in particular the finite grid instability. This numerical instability leads to numerical exponential growth of the total energy and destroys the simulations, when the grid spacing far exceeds the Debye length. In the simplest explicit PIC, this condition requires $\Delta x/\lambda_{De}<\pi$, a condition that can be somewhat relaxed using higher order interpolation. The ECsim eliminates this completely, allowing the grid spacing to be several orders of magnitude larger than the Debye length, we reported recently a case where the grid spacing was 16 orders of magnitude larger \citep{lapenta2016exactly}.  

Compared with fully implicit methods \citep{markidis2011energy, lapenta2011particle, chen2011energy} that also conserves energy exactly and is stable for any grid spacing, the ECsim differs for its linear formulation of the field equations that removes the need for the Newton non-linear iteration.

Compared with previous semi-implicit methods, such as the moment implicit \citep{brackbill-forslund,lapenta05,ipic3d} and the direct implicit \citep{directimplicit}, ECsim does not require any linearization step. The linearization used in the previous semi-implicit methods breaks energy conservation and $L_2$ stability, reintroducing the finite grid instability and a limitation to the size of the cells allowed. Practice and experience ensures that in the implicit moment method, the condition $\Delta x < \varsigma v_{th,e} \Delta t$ ensures no finite grid instability. This condition allows much larger cells than in the explicit PIC, but still prevents the method from exceeding the Debye scales by too large a factor. This factor, unfortunately is empirical and problem dependent. Practice shows that a plasma with low temperature electrons where the skin depth is much larger than the Debye length, $d_e>>\lambda_{De}$, is very hard to model without resolving very small scales. ECsim, instead, allows the user to completely ignore this limitation and set the grid spacing entirely on accuracy considerations without worrying about stability.

The results section highlights the advantages above in three practical cases. First, the ion acoustic wave shows the ability of ECsim  to resolve accurately the ion scales without needing to resolve the electron scales unnecessarily.  Second, the electron acoustic instability is used to show how ECsim can resolve one subpopulation of electrons without needing to resolve the smallest scales of the cold electrons. 
Finally, for a classic reconnection problem, the GEM challenge, ECsim shows the ability described above to resolve the electron scale at the skin depth level without suffering any finite grid instability even for very cold plasmas where $d_e>>\lambda_{De}$.

\section*{Acknowledgments}
The present work is supported by the US Air Force EOARD Project FA2386-14-1-0052, by the 
Onderzoekfonds KU Leuven (Research Fund KU Leuven, GOA scheme and Space Weaves RUN project) and by the
Interuniversity Attraction Poles Programme of the Belgian Science Policy Office (IAP
P7/08 CHARM). 
This research used resources of the National Energy Research
Scientific Computing Center, which is supported by the Office of
Science of the U.S. Department of Energy under Contract No. 
DE-AC02-05CH11231. Additional computing has been provided by NASA NAS and NCCS High Performance Computing, by the Flemish Supercomputing Center (VSC) and by PRACE Tier-0 allocations.


\begin{thebibliography}{33}
\expandafter\ifx\csname natexlab\endcsname\relax\def\natexlab#1{#1}\fi
\def\au#1{#1} \def\ed#1{#1} \def\yr#1{#1}\def\at#1{#1}\def\jt#1{\textit{#1}}
  \def\bt#1{#1}\def\bvol#1{\textbf{#1}} \def\vol#1{#1} \def\pg#1{#1}
  \def\publ#1{#1}\def\arxiv#1{#1}\def\org#1{#1}\def\st#1{\textit{#1}}

\bibitem[Balay {\em et~al.\/}(2016)Balay, Abhyankar, Adams, Brown, Brune,
  Buschelman, Dalcin, Eijkhout, Gropp, Kaushik, Knepley, McInnes, Rupp, Smith,
  Zampini, Zhang \& Zhang]{petsc-web-page}
{\sc \au{Balay, Satish}, \au{Abhyankar, Shrirang}, \au{Adams, Mark~F.},
  \au{Brown, Jed}, \au{Brune, Peter}, \au{Buschelman, Kris}, \au{Dalcin,
  Lisandro}, \au{Eijkhout, Victor}, \au{Gropp, William~D.}, \au{Kaushik,
  Dinesh}, \au{Knepley, Matthew~G.}, \au{McInnes, Lois~Curfman}, \au{Rupp,
  Karl}, \au{Smith, Barry~F.}, \au{Zampini, Stefano}, \au{Zhang, Hong} \&
  \au{Zhang, Hong}} \yr{2016} {PETS}c {W}eb page.
  {http://www.mcs.anl.gov/petsc}.

\bibitem[Birdsall \& Langdon(2004)]{birdsall-langdon}
{\sc \au{Birdsall, C.K.} \& \au{Langdon, A.B.}} \yr{2004} {\em Plasma Physics
  Via Computer Simulation\/}.  \publ{London: Taylor \& Francis}.

\bibitem[{Birn} {\em et~al.\/}(2001){Birn}, {Drake}, {Shay}, {Rogers},
  {Denton}, {Hesse}, {Kuznetsova}, {Ma}, {Bhattacharjee}, {Otto} \&
  {Pritchett}]{birnGEM}
{\sc \au{{Birn}, J.}, \au{{Drake}, J.~F.}, \au{{Shay}, M.~A.}, \au{{Rogers},
  B.~N.}, \au{{Denton}, R.~E.}, \au{{Hesse}, M.}, \au{{Kuznetsova}, M.},
  \au{{Ma}, Z.~W.}, \au{{Bhattacharjee}, A.}, \au{{Otto}, A.} \&
  \au{{Pritchett}, P.~L.}} \yr{2001}  \at{{Geospace Environmental Modeling
  (GEM) magnetic reconnection challenge}}.  \jt{JGR}  \bvol{106},
  \pg{3715--3720}.

\bibitem[{Birn} \& {Priest}(2007)]{birn-priest}
{\sc \au{{Birn}, J.} \& \au{{Priest}, E.~R.}} \yr{2007} {\em {Reconnection of
  magnetic fields: magnetohydrodynamics and collisionless theory and
  observations}\/}.  \publ{Cambridge University Press}.

\bibitem[{Biskamp}(2000)]{biskamp}
{\sc \au{{Biskamp}, D.}} \yr{2000} {\em {Magnetic Reconnection in Plasmas}\/}.
  \publ{Cambridge University Press, UK}.

\bibitem[Boor(1978)]{bspline}
{\sc \au{Boor, C.~De}} \yr{1978} {\em A practical guide to splines\/}.
  \publ{Springer}.

\bibitem[Brackbill \& Forslund(1982)]{brackbill-forslund}
{\sc \au{Brackbill, J.U.} \& \au{Forslund, D.W.}} \yr{1982}  \at{An implicit
  method for electromagnetic plasma simulation in two dimension}.  \jt{J.
  Comput. Phys.}  \bvol{46},  \pg{271}.

\bibitem[{Brackbill} \& {Cohen}(1985)]{Brackbill:1985}
{\sc \au{{Brackbill}, J.~U.} \& \au{{Cohen}, B.~I.}}, ed. \yr{1985} {\em
  {Multiple time scales.}\/}.

\bibitem[Burgess {\em et~al.\/}(1992)Burgess, Sulsky \&
  Brackbill]{burgess1992mass}
{\sc \au{Burgess, D}, \au{Sulsky, D} \& \au{Brackbill, JU}} \yr{1992}  \at{Mass
  matrix formulation of the flip particle-in-cell method}.  \jt{J. Comput.
  Phys.}  \bvol{103}~(1),  \pg{1--15}.

\bibitem[Chen {\em et~al.\/}(2011{\natexlab{{\em a\/}}})Chen, Chac\'on \&
  Barnes]{Chen-JCP-2011}
{\sc \au{Chen, G.}, \au{Chac\'on, L.} \& \au{Barnes, D.C.}}
  \yr{2011{\natexlab{{\em a\/}}}}  \at{An energy- and charge-conserving,
  implicit, electrostatic particle-in-cell algorithm}.  \jt{J. Comput. Phys.}
  \bvol{230},  \pg{7018}.

\bibitem[Chen {\em et~al.\/}(2011{\natexlab{{\em b\/}}})Chen, Chac{\'o}n \&
  Barnes]{chen2011energy}
{\sc \au{Chen, Guangye}, \au{Chac{\'o}n, Luis} \& \au{Barnes, Daniel~C}}
  \yr{2011{\natexlab{{\em b\/}}}}  \at{An energy-and charge-conserving,
  implicit, electrostatic particle-in-cell algorithm}.  \jt{J. Comput. Phys.}
  \bvol{230}~(18),  \pg{7018--7036}.

\bibitem[{Cohen} {\em et~al.\/}(1989){Cohen}, {Langdon}, {Hewett} \&
  {Procassini}]{directimplicit-optimisation}
{\sc \au{{Cohen}, B.~I.}, \au{{Langdon}, A.~B.}, \au{{Hewett}, D.~W.} \&
  \au{{Procassini}, R.~J.}} \yr{1989}  \at{{Performance and Optimization of
  Direct Implicit Particle Simulation}}.  \jt{J. Comput. Phys.}  \bvol{81},
  \pg{151--+}.

\bibitem[{Divin} {\em et~al.\/}(2012){Divin}, {Lapenta}, {Markidis}, {Newman}
  \& {Goldman}]{Divin2012}
{\sc \au{{Divin}, A.}, \au{{Lapenta}, G.}, \au{{Markidis}, S.}, \au{{Newman},
  D.~L.} \& \au{{Goldman}, M.~V.}} \yr{2012}  \at{{Numerical simulations of
  separatrix instabilities in collisionless magnetic reconnection}}.
  \jt{Physics of Plasmas}  \bvol{19}~(4),  \pg{042110}.

\bibitem[Divin {\em et~al.\/}(2010)Divin, Markidis, Lapenta, Semenov, Erkaev \&
  Biernat]{divin2010model}
{\sc \au{Divin, Andrey}, \au{Markidis, Stefano}, \au{Lapenta, Giovanni},
  \au{Semenov, VS}, \au{Erkaev, NV} \& \au{Biernat, HK}} \yr{2010}  \at{Model
  of electron pressure anisotropy in the electron diffusion region of
  collisionless magnetic reconnection}.  \jt{Physics of Plasmas (1994-present)}
   \bvol{17}~(12),  \pg{122102}.

\bibitem[Drake {\em et~al.\/}(2005)Drake, Shay, Thongthai \&
  Swisdak]{drake2005production}
{\sc \au{Drake, JF}, \au{Shay, MA}, \au{Thongthai, W} \& \au{Swisdak, M}}
  \yr{2005}  \at{Production of energetic electrons during magnetic
  reconnection}.  \jt{Physical Review Letters}  \bvol{94}~(9),  \pg{095001}.

\bibitem[{Fried} \& {Conte}(1961)]{Fried_and_Conte}
{\sc \au{{Fried}, B.~D.} \& \au{{Conte}, S.~P.}} \yr{1961} {\em The plasma
  dispersion function\/}.  \publ{Academic Press}.

\bibitem[{Gary}(2005)]{Gary_book}
{\sc \au{{Gary}, S.~P.}} \yr{2005} {\em Theory of space plasma
  microinstabilities\/}.  \publ{Cambridge University Press}.

\bibitem[Goldman {\em et~al.\/}(2011)Goldman, Lapenta, Newman, Markidis \&
  Che]{Goldman:2011}
{\sc \au{Goldman, M.~V.}, \au{Lapenta, G.}, \au{Newman, D.~L.}, \au{Markidis,
  S.} \& \au{Che, H.}} \yr{2011}  \at{Jet deflection by very weak guide fields
  during magnetic reconnection}.  \jt{Phys. Rev. Lett., accepted for
  publication} .

\bibitem[{Hewett} \& {Langdon}(1987)]{Hewett:1987}
{\sc \au{{Hewett}, D.~W.} \& \au{{Langdon}, A.~B.}} \yr{1987}
  \at{{Electromagnetic direct implicit plasma simulation}}.  \jt{J. Comput.
  Phys.}  \bvol{72},  \pg{121--155}.

\bibitem[Hockney \& Eastwood(1988)]{hockney-eastwood}
{\sc \au{Hockney, R.W.} \& \au{Eastwood, J.W.}} \yr{1988} {\em Computer
  simulation using particles\/}.  \publ{Taylor \& Francis}.

\bibitem[Langdon {\em et~al.\/}(1983)Langdon, Cohen \&
  Friedman]{directimplicit}
{\sc \au{Langdon, A.B.}, \au{Cohen, BI} \& \au{Friedman, A}} \yr{1983}
  \at{Direct implicit large time-step particle simulation of plasmas}.  \jt{J.
  Comput. Phys.}  \bvol{51},  \pg{107--138}.

\bibitem[Lapenta(2012)]{lapenta2012particle}
{\sc \au{Lapenta, Giovanni}} \yr{2012}  \at{Particle simulations of space
  weather}.  \jt{J. Comput. Phys.}  \bvol{231}~(3),  \pg{795--821}.

\bibitem[Lapenta(2016)]{lapenta2016exactly}
{\sc \au{Lapenta, Giovanni}} \yr{2016}  \at{Exactly energy conserving implicit
  moment particle in cell formulation}.  \jt{arXiv preprint arXiv:1602.06326} .

\bibitem[Lapenta {\em et~al.\/}(2006)Lapenta, Brackbill \& Ricci]{lapenta05}
{\sc \au{Lapenta, G.}, \au{Brackbill, J~.U.} \& \au{Ricci, P.}} \yr{2006}
  \at{Kinetic approach to microscopic-macroscopic coupling in space and
  laboratory plasmas}.  \jt{Phys. Plasmas}  \bvol{13},  \pg{055904}.

\bibitem[Lapenta \& Markidis(2011)]{lapenta2011particle}
{\sc \au{Lapenta, Giovanni} \& \au{Markidis, Stefano}} \yr{2011}  \at{Particle
  acceleration and energy conservation in particle in cell simulations}.
  \jt{Physics of Plasmas (1994-present)}  \bvol{18}~(7),  \pg{072101}.

\bibitem[{Lapenta} {\em et~al.\/}(2010){Lapenta}, {Markidis}, {Divin},
  {Goldman} \& {Newman}]{Lapenta2010}
{\sc \au{{Lapenta}, G.}, \au{{Markidis}, S.}, \au{{Divin}, A.}, \au{{Goldman},
  M.} \& \au{{Newman}, D.}} \yr{2010}  \at{{Scales of guide field reconnection
  at the hydrogen mass ratio}}.  \jt{Physics of Plasmas}  \bvol{17}~(8),
  \pg{082106}.

\bibitem[Lapenta {\em et~al.\/}(2011)Lapenta, Markidis, Divin, Goldman \&
  Newman]{lapenta2011bipolar}
{\sc \au{Lapenta, Giovanni}, \au{Markidis, Stefano}, \au{Divin, A},
  \au{Goldman, MV} \& \au{Newman, DL}} \yr{2011}  \at{Bipolar electric field
  signatures of reconnection separatrices for a hydrogen plasma at realistic
  guide fields}.  \jt{Geophysical Research Letters}  \bvol{38}~(17).

\bibitem[Lapenta {\em et~al.\/}(2014)Lapenta, Markidis, Divin, Newman \&
  Goldman]{lapenta2014separatrices}
{\sc \au{Lapenta, Giovanni}, \au{Markidis, Stefano}, \au{Divin, Andrey},
  \au{Newman, David} \& \au{Goldman, Martin}} \yr{2014}  \at{Separatrices: the
  crux of reconnection}.  \jt{Journal of Plasma Physics}  \pg{pp. 1--39}.

\bibitem[Markidis \& Lapenta(2011)]{markidis2011energy}
{\sc \au{Markidis, Stefano} \& \au{Lapenta, Giovanni}} \yr{2011}  \at{The
  energy conserving particle-in-cell method}.  \jt{J. Comput. Phys.}
  \bvol{230}~(18),  \pg{7037--7052}.

\bibitem[Markidis {\em et~al.\/}(2010)Markidis, Lapenta \&
  Rizwan-uddin]{ipic3d}
{\sc \au{Markidis, S.}, \au{Lapenta, G.} \& \au{Rizwan-uddin}} \yr{2010}
  \at{Multi-scale simulations of plasma with {iPIC3D}}.  \jt{Mathematics and
  Computers and Simulation}  \bvol{80},  \pg{1509--1519}.

\bibitem[Moses {\em et~al.\/}(1993)Moses, Finn \& Ling]{moses1993plasma}
{\sc \au{Moses, RW}, \au{Finn, JM} \& \au{Ling, KM}} \yr{1993}  \at{Plasma
  heating by collisionless magnetic reconnection: Analysis and computation}.
  \jt{Journal of Geophysical Research: Space Physics (1978--2012)}
  \bvol{98}~(A3),  \pg{4013--4040}.

\bibitem[Sulsky \& Brackbill(1991)]{sulsky1991numerical}
{\sc \au{Sulsky, Deborah} \& \au{Brackbill, JU}} \yr{1991}  \at{A numerical
  method for suspension flow}.  \jt{J. Comput. Phys.}  \bvol{96}~(2),
  \pg{339--368}.

\bibitem[Vu \& Brackbill(1992)]{vu}
{\sc \au{Vu, H.~X.} \& \au{Brackbill, J.~U.}} \yr{1992}  \at{Celest1d: An
  implicit, fully-kinetic model for low-frequency, electromagnetic plasma
  simulation}.  \jt{Comput. Phys. Comm.}  \bvol{69},  \pg{253}.

\end{thebibliography}

\end{document}